\input harvmac


%
\def\eql{~=~}

\def\coeff#1#2{\relax{\textstyle {#1 \over #2}}\displaystyle}
\def\half{{1 \over 2}}
\def\gop{\mathop{g}^{\!\circ}{}}
 \def\cB{{\cal B}}

 \def\cM{{\cal M}}
\def\cN{{\cal N}} \def\cO{{\cal O}}

\def\cV{{\cal V}}

\def\bfone{\relax{\rm 1\kern-.35em 1}}
\def\inbar{\vrule height1.55ex width.4pt depth0pt}
\def\IC{\relax\,\hbox{$\inbar\kern-.3em{\rm C}$}}
\def\ID{\relax{\rm I\kern-.18em D}}
\def\IF{\relax{\rm I\kern-.18em F}}
\def\IH{\relax{\rm I\kern-.18em H}}
\def\II{\relax{\rm I\kern-.17em I}}
\def\IN{\relax{\rm I\kern-.18em N}}
\def\IP{\relax{\rm I\kern-.18em P}}
\def\IQ{\relax\,\hbox{$\inbar\kern-.3em{\rm Q}$}}
\def\IR{\relax{\rm I\kern-.18em R}}
\def\us#1{\underline{#1}}
\font\cmss=cmss10 \font\cmsss=cmss10 at 7pt
\def\ZZ{\relax\ifmmode\mathchoice
{\hbox{\cmss Z\kern-.4em Z}} {\hbox{\cmss Z\kern-.4em Z}}
{\lower.9pt\hbox{\cmsss Z\kern-.4em Z}}
{\lower1.2pt\hbox{\cmsss Z\kern-.4em Z}}\else{\cmss Z\kern-.4em Z}\fi}

\def\gop{\mathop{g}^{\!\circ}{}}
\def\nihil#1{{\it #1}}
\def\eprt#1{{\tt #1}}
\def\nup#1({Nucl.\ Phys.\ $\us {B#1}$\ (}
\def\plt#1({Phys.\ Lett.\ $\us  {#1B}$\ (}
\def\cmp#1({Comm.\ Math.\ Phys.\ $\us  {#1}$\ (}
\def\prp#1({Phys.\ Rep.\ $\us  {#1}$\ (}
\def\prl#1({Phys.\ Rev.\ Lett.\ $\us  {#1}$\ (}
\def\prv#1({Phys.\ Rev.\ $\us  {#1}$\ (}
\def\mpl#1({Mod.\ Phys.\ Let.\ $\us  {A#1}$\ (}
\def\ijmp#1({Int.\ J.\ Mod.\ Phys.\ $\us{A#1}$\ (}
\def\atmp#1({Adv.\ Theor.\ Math.\ Phys.\ $\bf {#1}$\ (}
\def\jag#1({Jour.\ Alg.\ Geom.\ $\us {#1}$\ (}
\def\jhep#1({JHEP $\bf {#1}$\ (}

%


%
\lref\JMalda{J.~Maldacena, \nihil{The Large $N$ Limit of Superconformal
Field Theories and Supergravity,}, Adv.~Theor. Math. Phys.~{\bf 2}
(1998) 231 \eprt{hep-th/9711200}.}
%
\lref\GRW{M.\ G\"unaydin, L.J.\ Romans and N.P.\ Warner,
\nihil{Gauged $N=8$ Supergravity in Five Dimensions,}
Phys.~Lett.~{\bf 154B} (1985) 268; \nihil{Compact and Non-Compact
Gauged Supergravity Theories in Five Dimensions,}
\nup{272} (1986) 598.}
%
%
\lref\PolStr{J.~Polchinski and M.~J.~Strassler,
{\it The String Dual of a Confining Four-Dimensional Gauge Theory,}
\eprt{hep-th/0003136}.}
%
\lref\AKNW{A.~Khavaev and N.P.~Warner, \nihil{ A Class of ${\cal N} \! =\! 1$ 
Supersymmetric RG Flows  from  Five-dimensional ${\cal N} \! =\! 8$ 
Supergravity,} \plt{495} (2000) 215;  \eprt{hep-th/0009159}.}
%
\lref\KPW{A.\ Khavaev, K.\ Pilch and N.P.\ Warner, \nihil{New
Vacua of Gauged  ${\cal N}=8$ Supergravity in Five Dimensions},
\plt{487}  (2000) 14;  \eprt{hep-th/9812035}.}
%
\lref\FGPWa{D.~Z. Freedman, S.~S. Gubser, K.~Pilch, and N.~P. Warner,
\nihil{Renormalization Group Flows from Holography---Supersymmetry
and a c-Theorem,}   \atmp{3} (1999) 363; \eprt{hep-th/9904017} }
%
\lref\FGPWb{D.~Z. Freedman, S.~S. Gubser, K.~Pilch, and N.~P. Warner,
{\it Continuous Distribution of D3-branes and Gauged Supergravity,}
\jhep{7} (2000) 38; \eprt{hep-th/9906194}. }
%
\lref\KPNWa{K.\ Pilch and N.P.\ Warner, \nihil{A New Supersymmetric
Compactification of Chiral IIB Supergravity,} \plt{487} (2000) 22;
\eprt{hep-th/0002192}.}
%
\lref\KPNWb{K.~Pilch and N.P.~Warner, \nihil{$\cN=2$ Supersymmetric
RG Flows and the IIB  Dilaton,} \nup{594} (2001) 209;
\eprt{hep-th/0004063}.}
%
\lref\KPNWc{K.~Pilch and N.P.~Warner, \nihil{${\cal N} \! =\! 1$ 
Supersymmetric Renormalization  Group Flows from IIB Supergravity,} 
CITUSC/00-18, USC-00/02; \eprt{hep-th/0004063}.}
%
\lref\NWrevb{N.P.\ Warner, \nihil{Holographic Renormalization Group Flows: 
The View from Ten Dimensions,} Talk presented at the Second G\"ursey 
Memmorial Conference, to appear in the proceedings; \eprt{hep-th/0011207}.}
%
\lref\KLT{P.~Kraus, F.~Larsen and S.~P.~Trivedi,
\nihil{The Coulomb branch of gauge theory from rotating branes,}
\jhep{9903} (1999)   003; \eprt{hep-th/9811120}.}
%
\lref\CVJprobe{ C.~V.~Johnson, K.~J.~Lovis and D.~C.~Page,
\nihil{Probing some N = 1 AdS/CFT RG flows,}
\jhep{0105} (2001) 036;  \eprt{hep-th/0011166}.}
%
\lref\IBKS{I.~Bakas and K.~Sfetsos,
\nihil{States and curves of five-dimensional gauged supergravity,}
\nup{573} (2000)  768;  \eprt{hep-th/9909041}.}
%

\Title{ \vbox{ \hbox{CITUSC/01-021} \hbox{USC-01/03 } \hbox{\tt
hep-th/0106032} }} {\vbox{\vskip -1.0cm
\centerline{\hbox
{An ${\cal N} \!=\! 1$ Supersymmetric Coulomb Flow}}
\vskip 8 pt
\centerline{
\hbox{in $IIB$ Supergravity}}}}
\vskip -.3cm
\centerline{Alexei Khavaev and Nicholas P.\ Warner }
\medskip
\centerline{{\it Department of Physics and Astronomy}}
\centerline{{\it and}}
\centerline{{\it CIT-USC Center for
Theoretical Physics}}
\centerline{{\it University of Southern California}}
\centerline{{\it Los Angeles, CA 90089-0484, USA}}

\bigskip
\bigskip

We find a three-parameter family of solutions to 
$IIB$ supergravity that corresponds to $\cN=1$ supersymmetric
holographic  RG flows of $\cN=4$ supersymmetric Yang Mills theory.
This family of solutions allows one to give a mass to a
single chiral superfield, and to probe a two-dimensional
subspace of the Coulomb branch.   In particular, we examine
part of the Coulomb branch of the Leigh-Strassler fixed point.
We look at the infra-red asymptotics of these flows from the
ten-dimensional perspective.  We also make general conjectures
for the lifting Ansatz of five-dimensional scalar configurations to
ten-dimensional tensor gauge fields.  Our solution provides 
a highly non-trivial test of these conjectures.

\vskip .3in
\Date{\sl {June, 2001}}

\parskip=4pt plus 15pt minus 1pt
\baselineskip=15pt plus 2pt minus 1pt

\newsec{Introduction}

Holographic renormalization group flows have been
extensively studied via five dimension supergravity, but
have only been studied to a
lesser extent in the underlying ten-dimensional string theories
(or in $M$-theory).  It has become clear that the five-dimensional
descriptions of such flows can prove to be a powerful tool,
but that the ten-dimensional descriptions are essential to
a proper understanding of the infra-red behaviour of such flows
(see, for example, \NWrevb).

There are several issues in ``lifting'' five-dimensional
supergravity solutions to ten-dimensional supergravity.
First, is the obvious fact that a ten-dimensional lift may not
exist:  Many of the five-dimensional solutions that are considered
as holographic RG flows do not appear to be solutions of
a supergravity limit of an underlying string theory.  Thus,
while they may represent interesting conjectures, such flows may
have only a tenuous connection to the well-established
holographic string dualities.  There are, however, many five-dimensional
supergravity theories that are connected with an underlying
string compactification, and these  five-dimensional
supergravity theories fall into two  classes:  Effective
low energy theories, and, consistent truncations.  The former class
of solutions only represent approximations to some
higher-dimensional lifts, while the latter class of solutions  
will have  exact lifts to ten dimensions.  It is therefore the
consistent truncations that have the best chance of allowing
some reasonable interpretation of the infra-red end of the flow:
One only has to contend with the supergravity approximation
to string theory, and not with the possible breakdown of an effective
five-dimensional description.  This may suggest that knowing that
a five-dimensional theory is a consistent truncation  is
enough to extract all possible supergravity information
about the IR limit using that five-dimensional description.
This is, however, not true, and it has been shown in several
recent papers that it is essential to reconstruct the full
ten-dimensional solutions in order to understand properly the IR
asymptotics \refs{\PolStr, \KPNWb, \KPNWc, \NWrevb}.

In this letter we will consider gauged $\cN=8$
supergravity in five dimensions, which is widely believed
to a consistent truncation of the $S^5$ compactication of
$IIB$ supergravity.  The dual field theory is $\cN=4$ supersymmetric
Yang-Mills on $D3$-branes \JMalda.  The truncation to
gauged $\cN=8$ supergravity represents a truncation to
perturbations involving bilinear operators and
associated vevs in the Yang-Mills theory on the brane.
Several flows of this model have now been lifted to $IIB$
supergravity in ten dimensions \refs{\KPNWa,\KPNWb,\KPNWc}, but the general
story is far from complete.  The general formula for the
ten-dimensional metric of the lift is known \refs{\KPW,\KPNWb}, but 
the formulae for the lifts of the tensor gauge fields are not.
Our purpose here is two-fold.  First, we exhibit a new
ten-dimensional solution that represents a three-parameter
RG flow.  Two of the parameters represent independent scalar
masses or vevs, while the third represents a single fermion mass.
This enables us to extend the results of \refs{\KPNWc} that describe
the $\cN=1$ supersymmetric RG flow to the non-trivial
``Leigh-Strassler'' conformal fixed point.  In particular we are able to include
another Coulomb branch parameter and thus probe the Coulomb
branch of the Leigh-Strassler fixed point theory.  The five-dimensional
description of this three parameter family of flows was considered
in  \AKNW, but here we give the ten-dimensional lift,
and we are able to extract the brane geometry very explicitly.

The second purpose of this letter is to give
a conjecture for the general lift of all the tensor gauge fields
on the internal $5$-sphere in IIB supergravity.  This conjecture 
is based upon educated guess-work: It fits all known, non-trivial
lifts and provides the basic Ansatz for the solution
presented here.  Our new solution thus represents a highly
non-trivial test of the conjecture.

We will begin in section 2 by giving the general conjectured formulae
for lifting to ten dimensions.  In section 3 we will summarize
the results of \AKNW\ that are relevant, and we will
use this and our conjectures of section 2 to generate an
Ansatz for the new solution.  In section 4 we will give the
solution explicitly, and we will discuss its asymptotic behaviour.

\newsec{An Ansatz for Tensor Gauge Fields}

We will follow the conventions of \GRW\ throughout.
Recall the $42$ scalars of the $\cN=8$ theory in five dimensions
are parametrized by a $27 \times 27$ matrix of $E_{6(6)}$.
This matrix is naturally decomposed into blocks labelled:
$\cV_{J \alpha}{}^{a b}$ and $\cV^{IJ}{}^{a b}$, while
the corresponding blocks of the inverse matrix are labelled:
 ${\widetilde \cV^{J \alpha}}{}_{ab}$ and $\widetilde \cV_{IJ\,ab}$.
As was argued in \refs{\KPW,\KPNWb}, the full Ansatz for the
ten-dimensional metric is:
\eqn\dsmet{ds^2 ~=~ \Delta^{-{2 \over 3}}\, ds_{1,4}^2
~+~ d\hat s _5^2 \,,}
where the inverse metric of $d\hat s _5^2$ is given by:
\eqn\metanswer{\Delta^{-{2 \over 3}}\,\widehat g^{pq} \eql
 {1 \over a^2}\, K^{IJ\, p}\,K^{KL\, q}\, \widetilde
\cV_{IJab}\,\widetilde
\cV_{KLcd}\, \Omega^{ac}\,\Omega^{bd} \,.}
In this equation $K^{m IJ} = -K^{m JI}$, $I,J =1,\dots,6$ are
the Killing fields on the $S^5$, and $\Omega^{ab}$
is the $USp(8)$ symplectic invariant.   The quantity, $\Delta$,
is defined by
\eqn\Deltadefn{\Delta \,\equiv\,  \sqrt{{\rm det}(\hat g_{mp}\,
\gop^{pq})} \,,}
where the inverse  metric, $\displaystyle {\gop^{pq}}$, is that
of the ``round''  $S^5$.  The warp factor, $\Delta$, can thus
be determined by taking the determinant of both sides of
\metanswer.

The full Ansatz for the dilaton was proposed in  \KPNWb.
That is, the $IIB$ dilaton is represented by an $SL(2,\IR)$
matrix, $S$, and the gauge invariant quantity is the
matrix, $\cM =  S\,S^T$. This matrix, $\cM$,  appears in the kinetic
term of the $3$-form field strengths. It was argued in \KPNWb\ that the
dilaton Ansatz is given by:
\eqn\SAns{\Delta^{-{4 \over3}}\, \cM_{\alpha \beta} \eql
{\rm const}\times \cV_{I \alpha}{}^{ a b} \,\cV_{J \beta}{}^{ c d}
\,x^I x^J\, \Omega_{ac}\, \Omega_{bd} \,,}
where the $x^I$ are the cartesian coordinates for an
$\IR^6$ embedding of the compactification $5$-sphere.
That is, the $S^5$, and its deformations are defined
by the surface:  $\sum_I (x^I)^2 =1$.

The quantity, $\Delta$, can also be determined by taking the
determinant of both sides of \SAns.

The five-dimensional flows have a metric of
the form:
\eqn\fivemet{ds_{1,4}^2 ~=~ dr^2 ~+~ e^{2 A(r)}\big(
 \, \eta_{\mu \nu} \, dx^\mu \, d x^\nu \big) \,,}
In this letter we will use ``mostly +'' Lorentzian metrics.
A supersymmetric flow can usually be characterized in terms
of a superpotential, $W$, and this superpotential is
related to the scalar matrix as follows.  One defines
a $USp(8)$ tensor, $W_{ab}$, via:
\eqn\Wabdefn{W_{ab} ~\equiv~ - \epsilon^{\alpha\beta}
\,\delta^{IJ}\,\Omega^{cd} \,\cV_{I \alpha}{}_{ a c} 
\,\cV_{J \beta}{}_{b d} \,.}
One can often extract a superpotential from $W_{ab}$ when
the latter has a constant eigenvector.  More
precisely, the matrix  $W_{ac}W^{bc}$ is hermitian, and
symplectic invariance implies that the eigenvectors
come in symplectic pairs.  One can choose a $USp(8)$
gauge in which these eigenvectors are also
eigenvectors of $W_{ab}$, and if the eigenvectors
are constant then the eigenvalue, $W$, is often a 
superpotential \refs{\FGPWa,\AKNW}.   A  five-dimensional 
supersymmetric flow is then given by:
\eqn\floweqs{ {d \varphi_j \over d r} \eql {1 \over L}\,
{\del W \over \del \varphi_j}  \ , \qquad 
{d A \over d r}\eql  - {2 \over 3\,L}\,W \,, }
where $L$ is the radius of the $AdS_5$ at infinity, and
the $\varphi_j$ are canonically normalized scalars
with kinetic term $-\half \sum_j (\del \varphi_j)^2$.

The basic philosophy behind our conjectured Ansatz is to
identify the indices $I,J, \dots$ on $\cV$ as 
indices on $\IR^6$ and think of $x^I$ as the
unit normal to the deformed $S^5$  and then look
for building blocks that can be built out of
$\cV$.  The first step is to create a set of
{\it geometric} $W$-tensors, and in particular, define:
\eqn\Wtilde{\widetilde W_{ab} ~\equiv~ - \epsilon^{\alpha\beta}
\,x^I \, x^J \,\Omega^{cd} \,\cV_{I \alpha}{}_{ a c} 
\,\cV_{J \beta}{}_{b d} \,.}
In this definition we have replaced the $\IR^6$ metric,
$\delta^{IJ}$, in \Wabdefn\ by the outer product of
the unit normals, $x^I x^J$.  Now suppose that $\eta^a$
and $\zeta^a$ are two (constant) eigenvalues of $W_{ab}$ with
an eigenvalue $W$ that represents a superpotential.  Normalize
these vectors to have unit length, and consider:
\eqn\Wtildetrace{\widetilde W ~\equiv~ \half\, W_{ab} 
\big(\, \eta^a \, \eta^b ~+~ \zeta^a\, \zeta^b \,\big) \,.}
We will refer to $\widetilde W$ as the {\it geometric 
superpotential}.  Note that it contains more information
than just the superpotential, indeed the superpotential
can be obtained from $\widetilde W$ via:
$$
W ~=~  \half\, \sum_{I=1}^6 \, {\del^2 \over \del x^I \, \del x^I} 
\ \widetilde W \,.
$$

It is an empirical fact that in all the known lifts of 
flows to $IIB$ supergravity one has:
\eqn\AfourAns{A^{(4)}_{\mu \nu \rho \sigma} ~=~ 
\widetilde W \, e^{4 A(r)} \,  
\epsilon_{\mu \nu \rho \sigma} \,.}
Note that this expression is simply $\widetilde W$
times the volume form on the $D3$-brane measured
using the five-dimensional metric, \fivemet. While we have 
a heuristic justification of this formula, the primary 
argument in its favour is that it fits the all the lifted 
flows given in  \refs{\FGPWb,\KPNWa,\KPNWb,\KPNWc}.

In general, the Ansatz for the metric and other fields only
depends upon the matrix, $\cV$, whereas \AfourAns\ explicitly
depends upon the supersymmetry eigenvectors,  $\eta^a$ and 
$\zeta^a$, and implicitly upon a choice of superpotential.
Therefore we have only made an Ansatz appropriate to
supersymmetric flows.  On the other hand, it is 
possible that the particular form of \AfourAns\ is a convenient
gauge choice for $A^{(4)}$, and that there is a more
general formula for the field strength, $F^{(5)}$, in terms
of a geometric analog of the full supergravity potential.  Certainly
the radial derivative of \AfourAns\ generates precisely
the sort of terms one would need if such a conjecture were true.
There is also an obvious geometric generalization of the
tensor, $W_{abcd}$, of five-dimensional supergravity, and presumably
this would be a crucial ingredient in a geometrization of
the supergravity potential.
We will, however, not pursue this here since \AfourAns\ will
suffice for our purposes.

To give the Ansatz for the $2$-form gauge fields we need 
to introduce intrinsic coordinates, $\xi^j$, $j=1, \dots,5$ on the
$S^5$.  The partial derivatives, ${\del x^J \over \del \xi^j}$,
then act as projectors from $\IR^6$ onto $S^5$.
Consider the tensors:
\eqn\BAns{B^\alpha{}_{i j} ~=~ k \, L^2\, \cM^{\alpha \beta}
\big( x^K \, \cV_{K \beta}{}^{ a b} \big) \,
\Big(\cV_{IJab} \, {\del x^I \over \del \xi^i} \,
{\del x^J \over \del \xi^j}\Big) \,,}
where $\cM^{\alpha \beta}$ is the inverse of
$\cM_{\alpha \beta}$ defined in \SAns.  The constant, $k$,
is a dimensionless normalization constant.   For $\alpha=1,2$
this formula yields  $2$-forms on $S^5$, these fields
transform in an $SL(2,\IR)$ doublet, and at the linearized 
level the formula  produces  exactly the correct answer for 
the lowest modes of the $IIB$ supergravity $B$-fields.  
Indeed, we find that the foregoing formula exactly reproduces 
all of the $B$-fields (to all orders) for all of
the ten-dimensional lifted solutions obtained in \refs{\KPNWb,\KPNWc}
\foot{ There are some typographical errors in
\KPNWb\ and \KPNWc.  In \KPNWb\ the coefficient function $a_3$ for
the $\cN=2$ flow should have its sign reversed to give
$a_3 = -{4 \over g^2} {\sinh(2\,\chi) \over X_2} \sin\theta 
\cos^2 \theta$ , while for the $\cN=1$ flow in \KPNWc\ the 
coefficient function, $a_1$, should be multiplied by $i$ to 
give $a_1 = {2 i \over g^2} \tanh(\chi) \cos\theta$ .}.

We should stress that we have not proven that these formulae
are correct, we have merely constructed some moderately
obvious tensors from the five-dimensional scalar matrix, $\cV$, 
and we have checked that these formulae miraculously reproduce
some the rather complicated results of several known,
explicit solutions of $IIB$ supergravity. It is possible 
that these relatively simple formulae are a consequence
of the special subclasses of scalars that have been considered
in \refs{\FGPWb,\KPNWb,\KPNWc}, and that some modification will be needed 
in general. However, we will succumb to the obvious temptation, and conjecture
that \AfourAns\ and \BAns\ provide the exact lift of the
five-dimensional scalar fields to the tensor gauge fields.

\newsec{The $\cN=1$ Coulomb branch flows}

The flows that we consider are those that involve the
operators:
\eqn\scalops{\eqalign{
\cO_1 ~\equiv~ & \Tr (-X_1^2 - X_2^2 -X_3^2 - X_4^2 + 2\,
X_5^2 + 2\, X_6^2)  \,,\cr
\quad \cO_2 ~\equiv~ &\Tr (X_1^2 + X_2^2 - X_3^2 - X_4^2) \,,
\cr\cO_3 ~\equiv~ & \Tr (\lambda_4 \lambda_4) \, + \, h.c.  \,\,.}}
It should also be remembered that the operator:
\eqn\Ozero{\cO_0 ~\equiv~ \Tr \Big(\sum_{i=1}^6 X_i^2
\Big) \,,}
has no supergravity dual in the gauged $\cN=8$ supergravity theory,
but that the field theory on the brane always adds an appropriate
amount of $\cO_0$ to the operators $\cO_j, j=1,2$ so as to
preserve supersymmetry and positivity.   

We will denote the supergravity scalars dual to these operators
as $\alpha, \beta$ and $\chi$ respectively.  We introduce
$\rho \equiv e^\alpha, \nu \equiv e^\beta$, and define:
$\varphi_1 = {1 \over \sqrt{6}}\alpha, \varphi_2 = 
{1 \over \sqrt{2}}\beta$ and $\varphi_3= \chi$ and note that
the $\varphi_j$ are canonically normalized scalars with kinetic term
$-\half \sum_j (\del \varphi_j)^2$.

A superpotential for this flow was given in \AKNW: 
\eqn\Wpot{W ~=~ {1 \over 4 } \, \rho^4   \,
\big(\cosh(2\chi) - 3 \big)   ~-~  {1 \over 4 \,\rho^2 } \,
(\nu^2 + \nu^{-2} ) \, \big( \cosh(2\chi) +1 \big) \,  .}
We have replaced $\alpha\to -\alpha$ in the formula of
\AKNW\ so as to bring it into line with earlier papers
like \refs{\FGPWa,\KPNWb,\KPNWc}. The equations of motion are:
\eqn\floweqsred{ {d \alpha \over d r} \eql  {1 \over 6 L}\,
{\del W \over \del \alpha}  \ , \qquad {d \beta \over d r}
\eql { 1 \over 2 L}\,  {\del W \over \del \beta} \ ,\qquad
{d \chi \over d r} \eql   {1 \over L}\, {\del W \over \del
\chi}  \ .} 

It was shown in \AKNW\ that this superpotential describes a
family of $\cN=1$ supersymmetric flows  in which the chiral 
superfield, $\Phi_3$ is given a mass.  As is familiar from \FGPWa\
the flow of $\alpha$ and $\chi$ only describes a pure mass 
term for $\Phi_3$ for a specific choice of initial velocities.
The flow then runs to the non-trivial critical point of {\KPW}.
More generally, with other choices of the initial velocities,
and with $\beta \ne 0$ the flows can go to an unphysical 
regime ($\alpha \to -\infty$), or to  a 
region ($\alpha \to  +\infty$) in  which the flow approaches 
the Coulomb branch of the original $\cN=4$ region --
the fermion mass is swamped by the values of the vevs.  
In between these flows was an interesting pair
of ``ridge-line'' flows that started at the non-trivial fixed point. 
These flows correspond to Coulomb branch flows from
the Leigh-Strassler fixed point with non-zero vevs for
either the scalars in $\Tr(\Phi_1 \bar  \Phi_1)$ or 
$\Tr( \Phi_2 \bar \Phi_2)$.

The asymptotic behaviour of the supergravity scalars along these 
flows is:
\item{(i)} Unphysical Flow:
\eqn\alpmininf{\alpha \sim  \coeff{1}{20}\, \log(
\coeff{5}{3} \, r) \,, \quad \beta \sim \pm 3\,
\alpha \,,\quad  \chi  \sim   - 6\,\alpha \,, 
\quad A \sim 2\, \alpha
\sim \coeff{1}{10}\, \log (\coeff{5}{3} \, r) \,,}
\item{(ii)} Generic asymptotic flow to the $\cN=4$ Coulomb branch:
\eqn\alpposinf{ \chi \to a \,
r^{3 \over 4} \to 0\,, \quad
\alpha \sim  -\coeff{1}{4}\, \log\big(
\coeff{4}{3}\,  r \big)\,,
\quad \beta \to  \beta_0\,, \quad A \sim 
\coeff{1}{4}\, \log (r) \,,}
\item{(iii)} Ridge-line flow:
\eqn\ridgeflow{\eqalign{\alpha ~\sim~ & -{1 \over 4} \, \log(
\coeff{2}{3}\, r) \ , \quad  \beta ~\sim~ \pm( 3 \alpha
- \chi^2)\ , \cr
\chi^2 ~\sim~ & {1 \over a - 6\,   \log(r) } \ ,
\qquad A(r) ~\sim~ \log(r) \, .}}

It is useful to note that the flows with $\chi=0,\, \beta=0$
and  $\chi=0, \, \beta=\pm 3 \alpha$ represent three completely
equivalent $SO(4) \times SO(2)$ invariant Coulomb branch flows.  
These three different flows are simply discrete $SO(6)$ rotations
of one another.

The fermion mass parameter vanishes asymptotically
in both the physical flows, which is consistent with 
the dominance of a Coulomb parameters in the infra-red.  However,
the vanishing of $\chi$ occurs much more slowly
along the ridge-line flow.  Moreover, the 
five-dimensional geometry behaves very differently
because of the different asymptotics of $A(r)$.
To gain further insight into the holographic description
of these flows, we  will  examine and contrast them from
the ten-dimensional perspective.

\newsec{The solution to $IIB$ supergravity}

We now use the formulae of section 2 to obtain Ans\"atze
for the metric and tensor gauge fields for general values of
$\alpha, \beta$ and $\chi$.   We will parametrize the $5$-sphere
in $\IR^6$ by taking:
\eqn\varchng{\eqalign{ u_1 ~\equiv~& x_1+ i \, x_2 ~=~
\cos \theta \,\cos \phi \, e^{i\, \varphi_1} \,,  \quad
u_2 ~\equiv~ x_3 - i \, x_4 ~=~
\cos \theta \,\sin \phi \, e^{- i\, \varphi_2}\,, \cr
u_3 ~\equiv~& x_5 - i \, x_6 ~=~
\sin \theta  \, e^{- i\, \varphi_3} \,.}}
The ten-dimensional metric is then given by:
\eqn\metric{ds_{10}^2 ~=~ \Omega^2\,  ds_{1,4}^2 ~+~ ds_5^2\,,}
where $\Omega$, is given by:
\eqn\warpfac{\Omega  ~\equiv~ \Delta^{-{1 \over 3}}~=~ 
(\cosh\chi)^\half
\,\big( \rho^{-2} \,(\nu^2 \cos^2 \phi + \nu^{-2} \sin^2 \phi) \,
\cos^2 \theta ~+~ \rho^4 \, \sin^2 \theta\big)^{1 \over 4}\,,} 
and the metric $ds_5^2$ is the  following
metric on the deformed $S^5$:
\eqn\defSfive{\eqalign{ds_5^2 ~=~ & L^2\, \Omega ^{-2}\,
\Big[\rho^{-4} \, \big(\cos^2 \theta + \rho^6\, \sin^2 \theta \,
(\nu^{-2}   \cos^2\phi  +\nu^2\, \sin^2\phi) \big)\, d\theta^2 \cr
& + \rho^2 \cos^2 \theta \,(\nu^2 \,
\cos^2\phi  +\nu^{-2} \sin^2\phi) \, d\phi^2  \cr
&-2\,\rho^2 \, (\nu^2   - \nu^{-2} ) \,\sin\theta\, \cos \theta\,
\sin\phi\,\cos \phi\, d\theta \,d \phi \cr
&+~  \rho^2\, \cos^2 \theta \, (\nu^{-2}  
\cos^2\phi \, d\varphi_1^2  + \nu^2\, \sin^2\phi\, \, d\varphi_2^2) 
~+~ \rho^{-4} \, \sin^2 \theta \,  d\varphi_3^2 \,\Big]\cr
& +  L^2\,\Omega^{-6}\,\sinh^2 \chi\, \cosh^2 \chi \,\big(\cos^2\theta
\, (\cos^2 \phi \,  d\varphi_1 - \sin^2 \phi \,  d\varphi_2)  - 
\sin^2 \theta \,  d\varphi_3\,\big)^2 \,.}}
where $L$ is the radius of the round sphere.  

This form of the metric is very natural.  Recall that for $\chi=0$ this
metric must describe a Coulomb branch flow \refs{\FGPWb,\KLT},
and this in turn must be related to the extremal $D3$-brane 
solutions of \KLT.  Correcting a minor error in \KLT, and replacing
their radial coordinate by $\mu$, the metric of the extremal branes
is given by:
\eqn\rotbranes{\eqalign{ ds^2 \eql &  H^{-{1\over 2}}_{D3} \,
\big[ -dt^2 + dx^2_1 + dx^2_2 + dx^2_3\, \big] ~+~
 H^{1\over 2}_{D3} \, f^{-1}_{D3}\, {d\mu^2\over \prod_{i=1}^3
\big (1+{\ell^2_i\over  \mu^2}\big)}  \cr  
& + ~H^{1\over 2}_{D3}\, \mu^2\, \bigg[\bigg (1 + {\ell^2_1\, 
\cos^2\theta \over \mu^2} \,+\,  {\ell^2_2\, \sin^2\theta\, 
\sin^2\phi\over \mu^2} \,+\,  {\ell^2_3\, \sin^2\theta\, 
\cos^2\phi\over \mu^2}\bigg)\, d\theta^2 \cr
& +\, \Big (1 + {\ell^2_2\, \cos^2\phi
\over \mu^2} \,+\,  {\ell^2_3 \, \sin^2\phi \over \mu^2} \Big)\, 
\cos^2\theta\, d\phi^2 \cr
& -~ 2\, {\ell^2_2 - \ell^2_3\over \mu^2}\, 
\cos\theta\, \sin\theta\, \cos\phi
\, \sin\phi\,  d\theta \, d\phi + \Big(1+{\ell^2_1\over \mu^2}
\Big)\sin^2\theta d\varphi_1^2 \cr
&\qquad +~ \Big(1+{\ell^2_2\over \mu^2}\Big)\, \cos^2\theta\, 
\sin^2\phi\, d\varphi_2^2 + \Big(1+{\ell^2_3\over \mu^2}\Big)\, 
\cos^2\theta\, \cos^2\phi \,  d\varphi_3^2 \Big] \,.}}
where:
\eqn\Hfdefns{\eqalign{ H_{D3}  \eql & 1 + f_D {L^4\over \mu^4}\,, \cr
f_{D3}^{-1} \eql &\bigg ({\sin^2\theta\over 1 + {\ell^2_1\over \mu^2 }}
\, + \, {\cos^2\theta\sin^2\phi\over  1 + {\ell^2_2\over \mu^2 }} \, + \, 
{\cos^2\theta\cos^2\phi\over 1 + {\ell^2_3\over \mu^2 }}\bigg)\, 
\prod_{i=1}^3 ( 1 + {\ell^2_i\over \mu^2} ) \,.}}
As usual, in the near-brane limit where $L >> (\mu^2 + 
\ell_j^2)^\half$,  we ``drop the $1$'' in $H_{D3}$.

While \rotbranes\ apparently depends upon three parameters, this
is a fake in the near brane limit:  If $\ell_1=\ell_2=\ell_3$ then
the metric is still $AdS_5 \times S^5$, but merely written with
a non-standard radial coordinate.  It was a consequence of the
arguments in \FGPWb\ that this metric may then be mapped 
precisely onto \metric, and the detail of the mapping
 were given in \IBKS.  In particular, one has:
\eqn\maptorot{\eqalign{\rho ~=~ & \Big({p_2 \, p_3 \over p_1^2} 
\Big)^{1 \over 12}\,,\qquad  \nu ~=~ \Big({p_2   \over p_3 } 
\Big)^{1 \over 4}\,, \qquad \Omega ~=~ (p_1 \, p_2
\, p_3)^{-{1\over 6}} \, f_D^{-{1\over 4}} \,, \cr
e^{2\,A(r)} ~=~ & {\mu^2 \over L^2}\,(p_1 \, p_2
\, p_3)^{1 \over 3}  \,,\qquad {dr \over d \mu} ~=~ {L \over \mu}
\, (p_1 \, p_2 \, p_3)^{-{1\over 3}}\,; \qquad p_j ~\equiv~ 
\Big(1 + {\ell_j \over \mu^2} \Big)  \,.}}
The parameters, $\ell_j$ thus represent the initial data
of $\alpha$ and $\beta$ at infinity.

Observe that the metric \metric\ with $\chi \ne  0$ has 
the form of the Coulomb branch metric but with an
extra factor of $\cosh \chi$ in the warp factor.  Also
note that the last term in \defSfive\ may be written in terms
of the frame on the Hopf fiber:
\eqn\hopfing{L^2\, \Omega^{-6}\, \Big({\cal I}m \, \Big( u_1\,
d \bar u_1 + u_2\, d \bar u_2  +u_3\,d \bar u_3  \Big) \Big)^2\,.}
Thus the metric here is a straightforward generalization of that of 
\KPNWc:  The Coulomb branch metric is squashed by a factor
of $\cosh(\chi)$ while the Hopf fiber is stretched by a
factor proportional to $\tanh(\chi)$.

The geometric superpotential, $\widetilde W$, is given by evaluating
\Wtildetrace, and from this we obtain:
\eqn\Wtilde{\widetilde W ~=~ - \coeff{1}{8} \,  \rho^{-2} \,\big( 1 + 
\cosh (2\,\chi ) \big) \,  \cos^2 \theta \, \big(  \nu^2\, \cos^2  \phi 
+  \nu^{-2}\, \sin^2 \phi  \big)  ~+~  \coeff{1}{8} \, {\rho }^4  \,
\big( \cosh (2\,\chi )  -3\big) \, \sin^2 \theta \,.}

As noted in \AKNW, the dilaton and axion are constant upon
these flows, and so $\cM_{\alpha \beta} = \delta_{\alpha \beta}$.

The Ansatz, \BAns\ yields the following results for the tensor
gauge fields.  Let $\cB_{\mu \nu} = B^1_{\mu \nu} + i B^2_{\mu \nu}$,
then:
\eqn\CalcBfield{\eqalign{ \cB ~\equiv~ & \coeff{1}{2} \, 
 \cB_{\mu \nu}  \, dx^\mu \wedge dx^\nu \cr
\eql & \coeff{1}{2}\,\Omega^{-4} \, L^2\, \sinh(2\chi)\, \big(\rho^4 
\, u_3 \, du_1 
\wedge du_2 + \rho^{-2}\, \nu^2  \, u_1 \, du_2 \wedge du_3 + 
\rho^{-2}\, \nu^{-2} \, u_2 \, du_3 \wedge du_1 \big) \,.}}

It is a tedious, but straightforward exercise to verify that
\metric, \AfourAns\ with \Wtilde\ and \CalcBfield,  do indeed 
satisfy the equations of motion of $IIB$ supergravity.
To verify this one must, of course, use the equations of
motion, \floweqs\ and \floweqsred, with the superpotential
\Wpot.

Thus our conjectured general consistent truncation Ansatz has
passed a very non-trivial test, and we have a three parameter
family of holographic RG flows in ten dimensions.

\newsec{Infra-red limits and brane probes}

The supergravity solution presented here contains the $\cN=1$ flows
already discussed in \KPNWc.  Indeed, one can verify that if
one sets $\beta=0$, or $\nu=1$, in all the equations in the 
previous section, one does indeed recover the solution of \KPNWc.
We will therefore not dwell
upon these aspects of our solution, but consider the new aspects
associated with the flows in $\beta$.

The physical flows identified in section 3 both have $\chi \to 0$,
but at very different rates.  It is evident from \metric, \warpfac\ 
and \defSfive\ that if $\chi\to 0$ then the metric limits directly to
the extremal rotating brane metric associated with the Coulomb branch
of the $\cN=4$ theory \refs{\KLT,\FGPWb}.   Thus, in the infra-red, all
of these $\cN=1$ Coulomb branch flows approach the $\cN=4$ Coulomb
branch flows, which suggests that the mass that has been turned on for
$\Phi_3$ is ultimately swamped by the Coulomb vevs, and the model retains 
knowledge of its $\cN=4$ structure.  

To understand the roles of generic flows, (ii), and  the ridge-line 
flows, (iii), it is instructive  to consider their detailed asymptotics.
Recall that the $SO(2) \times SO(2) \times SO(2)$ invariant Coulomb branch 
flows ($\chi=0$) are sourced by  an ellipsoidal distribution of $D3$-branes,
with semi-major axes determined by the $\ell_j$.  Indeed, $\ell_1$ is 
the semi-major axis in the $(x_5,x_6)$ direction, while $\ell_2$ is 
the semi-major axis in the $(x_3,x_4)$ direction and $\ell_3$ is 
the semi-major axis in the $(x_1,x_2)$ direction\foot{To make this
correspondence more precise one must make a change of variables to
coordinates, $y_a$, defined in \KLT.}.

Using the correspondence \maptorot\ it is thus easy to determine
the asymptotic limits of the physical flows:
\item{a)}  If $\rho \to \infty$, $\nu \to \nu_0$ then $\ell_1 =0$,
but $\nu_0^2 = {\ell_2  \over \ell_3}$.  The distribution of branes
is thus an ellipsoidal shell in $(x_1,x_2,x_3,x_4)$ with a 
$\delta$-function in the other two directions.  In this limit one
also has:
$$
\cB ~\sim~   \coeff{1}{2}\,\Omega^{-4} \, L^2\,  \rho^4 \,  \sinh(2\chi)\, 
 \, u_3 \, du_1  \wedge du_2   \,.
$$
Thus $B^{NS}$ and $B^{RR}$ are both parallel to the brane distribution.
\item{b)}  If $\rho \to \infty$, $\nu \sim \rho^{-3}$ then 
$\ell_1 = \ell_2 =0$, and the distribution of branes
is a disk in the $(x_1,x_2)$  direction.  In this limit one
also has:
$$
\cB ~\sim~   \coeff{1}{2}\,\Omega^{-4} \, L^2\, \rho^4 
\, \sinh(2\chi) \,  du_1 \wedge \big( u_3 \, du_2 ~-~  
 u_2 \, du_3  \big) \,.
$$
Thus $B^{NS}$ and $B^{RR}$ each have  ``a leg'' in the 
brane distribution, and a leg perpendicular to it.
\item{c)}  The flow with $\rho \to \infty$, $\nu \sim \rho^3$
is the same as in b), but with $\ell_1 = \ell_3 =0$ and with 
$u_1$ and $u_2$ interchanged.

The foregoing asymptotic behaviour is completely
consistent with  the field theory interpretation proposed in \AKNW.
First, the superfield, $\Phi_3$ has been given a mass, and
so the only remaining fields that can receive vevs are 
$\Phi_1$ and $\Phi_2$, and these correspond to spreading the
branes in the $u_1$ and $u_2$ directions respectively.  The ``generic'' 
flow  limits to a two parameter family of flows that reflect the 
initial conditions of $\alpha$ and $\beta$, and correspond to
the different possible scales of the vevs in $\Phi_1$ and
$\Phi_2$.  Thus the ridge-line flows emerge as natural boundaries of the
``generic'' flows: either $\ell_2$ or $\ell_3$ vanishes, collapsing
the ellipsoidal shell to a disk.  The ``generic'' flow towards the $\cN=4$ 
Coulomb branch also washes out the $\cB$-field much more rapidly than the 
ridge-line flow.   

The foregoing picture focusses closely upon the 
Coulomb branch structure of the theory.  It should be remembered
that there is a non-trivial critical  point corresponding
to the Leigh-Strassler fixed point theory.  This fixed point 
theory is conformal, and at the fixed point, the field, $\Phi_3$,
has been ``integrated out.''   Moreover, while there are 
certainly pure Coulomb branch flows with $\beta \sim \pm 3 \alpha$,
the ridge-line flow, \ridgeflow\ only seems accessible from this
non-trivial fixed point.  More general physical flows from this
fixed point are of the form \alpposinf\ and they wash out the 
$\cB$-field much more rapidly.  We find it intriguing that 
the Coulomb branch flows of the Leigh-Strassler point with 
$\Phi_1,\Phi_2$ both non-zero rapidly flow towards the 
$\cN=4$ Coulomb branch, but that the flows with only $\Phi_1 
\ne 0$ or  $\Phi_2  \ne 0$ appear to be privileged 
in that the $\cB$ field vanishes far more slowly.  It would
be very interesting to understand this more deeply from
the perspective of the physics on the brane.
 
Finally, we have also performed the brane probe calculation for the
supergravity solution presented here, and the results 
do not differ significantly from those of \CVJprobe.   
This is not very surprising since we are generalizing the
result by adding another Coulomb branch parameter.  
We find that  potential felt by the brane probes is given by:
\eqn\probpot{ V ~=~   e^{4\, A(r)} \, \big(\Omega^4 ~-~ 4 \, 
\widetilde W \big) ~=~ e^{4\, A(r)}  \, \rho^4\, (\cosh(2 \, \chi)
- 1)\, \sin^2 \theta \,,}
which is exactly the same as was found in \CVJprobe. 
The potential vanishes for $\theta =0$, and $D3$-brane probes
have the following metric on the $4$-dimensional moduli space 
transverse to the branes:
\eqn\metonmod{\eqalign{ds^2 ~=~ & \coeff{1}{2}\, \tau_3 \, e^{2\,A} \,
\big[ \zeta \, (  \rho^{-2} \cosh^2 (\chi) \, dr^2
~+~ L^2\, \rho^2 \, d \phi^2 ) ~+~   L^2\, \rho^2 \,(\nu^{-2} \cos^2 \phi \,
d\varphi_1^2 \cr & 
+ \nu^2 \sin^2 \phi \, d\varphi_2^2)~+~  L^2\, \rho^2 \, 
\sinh^2(\chi) \, \zeta^{-1}\, (  \cos^2 \phi \, d\varphi_1  - 
 \sin^2 \phi \, d\varphi_2 )^2  \big]  \,,}}
where
\eqn\zetadef{ \zeta ~\equiv~ (\nu^2\, \cos^2 \phi ~+~ 
\nu^{-2} \, \sin^2 \phi ) \,.}
This is essentially an ellipsoidally squashed version of the metric
obtained in \CVJprobe.

\bigskip
\leftline{\bf Acknowledgements}
We would like to thank K.~Pilch for helpful conversations.
This work was supported in part by funds provided by the DOE
under grant number DE-FG03-84ER-40168.

\listrefs
\vfill
\eject
\end